\documentclass{svjour2}                    
\smartqed  
\usepackage{graphicx}
\def\d{\partial}
\def\vr{\vec{r}}
\def\tv{\tilde{V}}
\def\tta{\tilde{\theta}}
\def\k{\vec{k}}

\def\dep{\delta\psi}
\def\tdp{\tilde{\delta\psi}}
\def\rtdr{\frac{\tilde{\delta\rho}}{\rho_o}}
\def\rdr{\frac{\delta\rho}{\rho_o}}
\def\dr{\delta\rho}
\def\cU{\mathcal{U}}
\def\tU{\tilde{U}}

\begin{document}

\title{Static spectroscopy of a dense superfluid}

\author{S. Villerot \and B. Castaing \and L. Chevillard}

\institute{Laboratoire de Physique de l'Ecole normale sup\'erieure de Lyon\\ CNRS, Universit\'e de Lyon - 46 all\'ee d'Italie, 69364 Lyon Cedex 07, France}

\titlerunning{Static spectroscopy of a dense superfluid}        

\authorrunning{S. Villerot \etal} 

\date{Received: date / Accepted: date}

\maketitle

\begin{abstract}
Dense Bose superfluids, as HeII, differ from dilute ones by the existence of a roton minimum in their excitation spectrum. It is known that this roton minimum is qualitatively responsible for density oscillations close to any singularity, such as vortex cores, or close to solid boundaries. We show that the period of these oscillations, and their exponential decrease with the distance to the singularity, are fully determined by the position and the width of the roton minimum. Only an overall amplitude factor and a phase shift are shown to depend on the details of the interaction potential. Reciprocally, it allows for determining the characteristics of this roton minimum from static ``observations'' of a disturbed ground state, in cases where the dynamics is not easily accessible. We focus on the vortex example. Our analysis further shows why the energy of these oscillations is negligible compared to the kinetic energy, which limits their influence on the vortex dynamics, except for high curvatures.
\end{abstract}

\section{Introduction}
\label{intro}
HeII is the low temperature, low pressure superfluid phase of $^{4}$He. The roton minimum, in its excitation spectrum, 
has been inferred by Landau \cite{Landau} from the viscosity measurements of Andronikashvili \cite{Andro}. It has been shown, 
by Feynman \cite{Feynman} to be due to the dense packing of $^{4}$He atoms. Solidification, which occurs above 2.5MPa 
for low temperature $^{4}$He, can be seen as a condensation of rotons, due to their interactions.

The low value of the excitation energy at the roton minimum also suggests that the superfluid has a strong susceptibility 
for spatial perturbations of wave number $k_{o}$, the position of the roton minimum. Localized perturbations will then produce oscillations in the superfluid density in their neighborhood \cite{Dal92,Pomeau,Berloff}.  This has been known for long. For instance, using an ansatz for the superfluid order parameter, based on the only neighborhood of the roton minimum, T. Regge \cite{Regge} could show the existence of these oscillations close to the free surface of superfluid $^4$He. He could obtain the surface tension within 30\% of the experimental value. More recent theoretical works \cite{Dal92,Berloff} obtain such oscillations in the order parameter close to other singularities, as vortex cores. However, the sensitivity of these oscillations to details of the real problem at hand is not clear. 

Up to now, because of the smallness of the scales involved in this phenomenon (e.g. of the order of the Angstrom in $^4$He), any experimental investigations remain a great challenge. For instance, one could think of neutron diffraction by a vortex lattice, created in a $^4$He film by a rapid rotation. However, rotation frequencies as large as 10$^5$rd/s would result in a distance between vortices of order 1 $\mu$m, which ask for resolving orders up to 10$^3$ in the reciprocal lattice

So then, two approaches can be followed. First ``numerical experiments'' through {\it ab initio} calculations \cite{Reatto2} allow to compare the ground state and the one vortex state for a limited number of atoms. Secondly, Gross-Pitaevskii equations are often proposed for modeling the dynamics, and space dependence of the order parameter. The interaction term in these equations is chosen so as to fit the dispersion curve of elementary excitations. But a large class of such interaction terms can fit the same dispersion curve. What is the influence of their differences? What is the influence of details of the dispersion curve such as the phonon (long wavelength), maxons (maximum of the dispersion curve) or very short wavelength regions? Up to what precision do we have to fit the roton minimum (simple parabola, skewness)? At the end, it is known that, for strongly interacting superfluids as $^4$He, density could not be simply proportional to the squared order parameter, as in dilute systems.

Determining the parameters which control the extension and amplitude of these oscillations, and discussing their implications are the goals of this paper. We shall focus on density oscillations far from the singularity, where they are small and amenable to a linear equation. We shall compare our approximation both with ``exact'' resolution of various Gross-Pitaevskii models, and with the results of {\it ab initio} calculations.

The existence of density oscillations close to a vortex core is expected to have tremendous importance in the study of quantum turbulence in which a central role is played by vortex reconnections \cite{BarDon01,VinNie02} and thus it is useful to quantify how these oscillations 
depend on the shape of the roton minimum. In particular, this study could be necessary to the design of numerical investigations of vortex reconnections, as it was done formerly without taking into account the roton gap \cite{KopLev93}. It is why we shall focus on the vortex problem.

This paper is organized as follows. First and foremost, in section \ref{pr}, the universal influence of the roton minimum close to a localized perturbation is shown on a simple example, the Pomeau-Rica version \cite{Pomeau} of the Gross-Pitaevskii equation \cite{Stringari}. In the following section \ref{vor}, we extend this linear approximation to density oscillations in real superfluids. The order parameter oscillations around a vortex core are then compared, section \ref{numpr}, in the Pomeau-Rica model, to our linear approximation. Then, in section \ref{He}, we apply the same comparison to a more elaborated Gross-Pitaevskii model \cite{Berloff}, aiming at better reproducing the full dispersion curve than in the Pomeau-Rica version. Our last comparison, section \ref{reatto}, concerns the density profile around a vortex obtained {\it via ab initio} variational calculations \cite{Reatto2}. The last section, \ref{eng}, is devoted to energy considerations. We finally conclude on the implications for real systems.

\section{Linear treatment of the non-local Gross-Pitaevskii equation}
\label{pr}

The Pomeau-Rica \cite{Pomeau} version of the Gross-Pitaevskii \cite{Stringari} equation reads:
\begin{equation}
 i\hbar\d_t\psi(\vr\!,\!t)\!=\!-\frac{\hbar^2}{2m}\Delta\psi(\vr\!,\!t)\! -\!\mu\psi(\vr\!,\!t)   
\!+\! U_o\psi(\vr\!,\!t)\!\!\int\!\! d^3\vr'\,\theta(|\vr\!-\!\vr'|/a)|\psi(\vr'\!,\!t)|^2
\label{pori}
\end{equation}
where  $U_o\theta(|\vr-\vr'|/a)$ represents the interaction potential between two atoms located at $\vr$ and $\vr'$, 
$a$ is the range of the potential. The interaction $\theta(\vec{x})$ is 1 if $|\vec{x}|<1$, 0 for $|\vec{x}|>1$ and 
$\mu$ is the chemical potential, fixing the equilibrium density $n=|\psi(\vr,t)|^2$.

Using the transformation $\vr \rightarrow a\vr$, $\psi \rightarrow \sqrt{n}\psi$, $t \rightarrow (2ma^2/\hbar)t$, 
Eq. \ref{pori} can be written in the following non-dimensional form:
\begin{equation}
 i\d_t\psi(\vr,t)\!=\!-\!\Delta\psi(\vr,t)\! -\!4\pi\Lambda\psi(\vr,t)/3\! +\!\Lambda\psi(\vr,t)\!\int \!d^3\vr'\,\theta(|\vr-\vr'|)|\psi(\vr',t)|^2
\label{ndpr}
\end{equation}
where
\begin{equation}
 \Lambda=\frac{2ma^2U_ona^3}{\hbar^2}
\label{lambda}
\end{equation}
remains the only parameter. Any static perturbation will appear as an additional term $V(\vr)=\psi(\vr)\mathcal V(\vr)$ in the right hand side of 
equation (\ref{ndpr}). Far from the perturbation, the solution of Eq. \ref{ndpr} can be written to linear order as 
$\psi=1+\dep$, with $\dep$ and its complex conjugate $\dep^*$ verifying the linear system 
\begin{eqnarray}
 i\d_t\dep(\vr,t)+\Delta\dep(\vr,t) \mbox{\hspace{45mm}}  \nonumber \\ 
- \Lambda\int d^3\vr'\,\theta(|\vr-\vr'|)(\dep(\vr',t)+\dep^*(\vr',t))=V(\vr)  \\
 -i\d_t\dep^*(\vr,t)+\Delta\dep^*(\vr,t)  \mbox{\hspace{40mm}} \nonumber \\ 
- \Lambda\int d^3\vr'\,\theta(|\vr-\vr'|)(\dep(\vr',t)+\dep^*(\vr',t))=V(\vr)
\label{dpsi}
\end{eqnarray}
By Fourier transforming, we get:
\begin{eqnarray}
(\omega-k^2- \Lambda\tta(k))\tdp(\k,\omega)- \Lambda\tta(k)\tdp^*(\k,\omega)=\tv(\k) \nonumber\\   
(\omega+k^2+ \Lambda\tta(k))\tdp^*(\k,\omega)+ \Lambda\tta(k)\tdp(\k,\omega)=-\tv(\k)  
\label{dpf}
\end{eqnarray}
where $\tta(k)$ is the Fourier transform of the normalized interaction potential $\theta$:
\begin{equation}
 \tta(k)=\frac{4\pi}{k^3}(\sin(k)-k\cos(k))
\end{equation}
Eliminating $\tdp^*$ from the former system (Eq. \ref{dpf}) gives:
\begin{equation}
 (\omega^2-\omega(k)^2)\tdp(\k,\omega)=(\omega+k^2)\tv(\k)
\label{tdp}
\end{equation}
where $\omega(k)$ is the dispersion relation of the linear excitations, namely
\begin{equation}
\omega(k)^2=k^2(k^2+ 2\Lambda\tta(k))\mbox{ .}
\end{equation}
The frequency $\omega = \omega(k)$ allows a non zero $\tdp(\k,\omega)$ without any perturbation ($\tv(\k)=0$). 
The static ($\omega=0$) solution of equation 
(\ref{tdp}) is:
\begin{equation}
\tdp(\k)=-\frac{k^2\tv(\k)}{\omega(k)^2}\mbox{ .}
\label{tdp2}
\end{equation}

In this model, $\tdp(\k)$ represents half the Fourier transform of the relative density deformation:
\begin{equation}
\rtdr(\k)=-\frac{2k^2\tv(\k)}{\omega(k)^2}\mbox{ .}
\label{tdr}
\end{equation}

Finally, the form of the deformation $\dr(\vr)/\rho_o$ in the physical space is given by the inverse Fourier transform
\begin{equation}\label{eq:InvFT}
 \rdr(\vr) = \frac{1}{(2\pi)^3}\int \rtdr(\k) e^{i\k.\vr}d\k\mbox{ .}
\end{equation}

In the case of interest for us, $\omega(k)^2$ has a deep minimum (the roton minimum) at $k=k_o$ that will clearly dominate the 
right hand side (RHS) of Eq. \ref{tdp2}. Let us write in its 
neighborhood $\omega(k)^2=\Omega^2+c^2(k-k_o)^2$, where  $\hbar\Omega$ is the energy roton gap and 
$m_*=\hbar\Omega/c^2$ the ``roton mass''.

\section{Extension to real systems. Linear vortex}
\label{vor}

Equation (\ref{tdr}) is central in our paper. Let us present here our arguments for extending it to any real superfluid, as $^4$He. We are interested in the deformation $\dr(\vr)$ far from the perturbation. As $\dr/\rho_o$ is small, it should obey a linear equation:
\begin{equation}
 \mathcal{L}(\rdr)=V(\vr,t)
\end{equation}

Due to space and time translational invariance, the Fourier transform of this equation is local:
\begin{equation}
 \tilde{F}(\omega,\k)(\omega^2-\omega(\k)^2)\rtdr(\omega,\k)=\tilde{V}(\omega,\k)
\end{equation}

The shape we gave to this equation simply expresses that deformations can propagate in the superfluid, in the absence of any perturbation, if $\omega=\pm \omega(\k)$. Our point is that $\tilde{F}$ is smooth in the neighborhood of $k=k_o$, as well as $\tilde{V}$, if the perturbation is localized. For a static perturbation, we can then write, in the general case:
\begin{equation}
 \rtdr(\k)=-\frac{\tv(\k)/\tilde{F}(\k)}{\omega(k)^2}\mbox{ .}
\label{tdr2}
\end{equation}

For a plane localized perturbation, as a free surface, or a wall, we can write $\tv(\k)/\tilde{F}(\k)$ $= \tilde{v}(k_x)\delta(k_y)\delta(k_z)$, where $\delta$ is the Dirac function, and the plane is perpendicular to the $x$ axis.  Because the roton minimum is assumed deep, i.e. $\Omega^2$ is assumed small, the RHS of Eq. \ref{tdr2} has a sharp maximum at $k=k_o$. Thus, the very precise shape of the perturbation is not important as long as it is localized, i.e. the Fourier transform $\tilde{v}(k_x)$ is smooth, and we can take it as a constant $\tilde{v}(k_o)$. The result:
\begin{equation}
 \rtdr(\k)=-\frac{\tilde{v}(k_o)}{\omega(k)^2}\mbox{ ,}
\label{tdr3}
\end{equation}
is similar to the ansatz proposed by Regge \cite{Regge}.

Let us now turn to the linear vortex case.

To model the perturbation induced by a vortex line, we consider a 
localized perturbation  that depends only on the distance $r$ from the $z$-axis\footnote{Doing so, we neglect the influence of the phase, which changes by 
$2\pi$ when turning around the vortex.} (cylindrical symmetry). In this case, we can write $\tv(\k)/\tilde{F}(\k) = \tilde{v}(k_{\perp})\delta(k_z)$, where $\delta$ is the Dirac function.  
Again, due to the sharpness of $1/\omega(k)^2$ close to the roton minimum, the very precise shape of the perturbation is not important as long as it is localized, i.e. the Fourier 
transform $\tilde{v}(k_{\perp})$ is smooth. In this spirit, we can Taylor expand the perturbation as $\tilde{v}(k_{\perp})
\approx (\tilde{v}(k_o) + \tilde{v}'(k_{o}) (k_{\perp}-k_o))\delta(k_z)$.
Using Eq. \ref{tdr3}, we can approximate $\rtdr(\k)$ as:
\begin{equation}
 \rtdr(\k)\approx -\frac{\tilde{v}(k_o) + \tilde{v}'(k_{o}) (k_{\perp}-k_o)}{\Omega^2+c^2(k-k_o)^2}\delta(k_z)\mbox{ .}
\label{tdr4}
\end{equation}


Then, using the approximative form of $\rtdr(\k)$ (Eq. \ref{tdr4}) and Eq. \ref{eq:InvFT} , we get $\dr(\vr)= \dr(r)$ with
\begin{equation}\label{eq:FormPertExact}
\rdr(r)\approx \frac{1}{2\pi}
\int_0^{\infty}\frac{\tilde{v}(k_o) + \tilde{v}'(k_{o}) (k_{\perp}-k_o)}{\Omega^2+c^2(k_{\perp}-k_o)^2}J_0(k_{\perp}r)k_{\perp}dk_{\perp}\mbox{ ,}
\end{equation}
where $J_0(x)=(1/2\pi)\int_0^{2\pi}\exp(ix\cos\phi)d\phi$ is the Bessel function of the first kind. The shape of the perturbation (Eq. \ref{eq:FormPertExact}) can be accurately numerically evaluated and is found oscillatory (data not shown). Given the approximative form used in Eq. \ref{tdr4}, it is expected to be valid only when it is small, {\it i.e.} for $r$ relatively large. In the integral, we can thus consider $k_{\perp}r$ as large, for the interesting $k_{\perp}$ values. This allows to use the following asymptotic form of the Bessel function
\begin{equation}\label{eq:AsymptBessel}
 J_0(x)\approx \frac{2\cos(x-\pi/4)}{\sqrt{2\pi x}}\mbox{ .}
\end{equation}
Using the asymptotic form of the Bessel function (Eq. \ref{eq:AsymptBessel}), the $k_\perp$ integration entering in 
Eq. \ref{eq:FormPertExact} can be performed in the complex plane, which gives:
\begin{equation}
 \rdr(r)  \approx \sigma\frac{\exp(-k_1r)}{\sqrt{k_or}}\cos\left(k_or-\pi/4 + \frac{k_1}{2k_o}+\phi_o\right)
 \label{foa}
\end{equation}
with $k_1=\Omega/c$ and:
\begin{eqnarray}
 \sigma & = & \frac{\gamma}{(2\pi)^{1/2}}\frac{k_o}{c^2k_1} \nonumber \\
 \gamma ^2 & = & \tilde{v}(k_o)^2 +k_1^2 \tilde{v}'(k_{o})^2  \nonumber \\
 \cos(\phi_o) & = & \frac{\tilde{v}(k_o)}{\gamma}
 \label{foa2}
\end{eqnarray}
the sign of $\gamma$ being chosen such that $-\frac{\pi}{2}<\phi_o<\frac{\pi}{2}$.

Thanks to Eq. \ref{foa}, we can see that the density profile close to the 
vortex behaves as an oscillatory function of wavelength $2\pi/k_o$ that exponentially tends to zero with a characteristic length scale $1/k_1$. 
Let us emphasize that the obtained shape of the deformation $\dr(r)$ 
only depends of $k_or$, $k_1/k_o$ and an additional phase-shift $\phi_o$. In the small $\Omega$ limit (i.e. deep minimum), 
the ratio $k_1/k_o$ is expected to be small 
compared to unity. The sign and amplitude of $\phi_o$ depends on the shape and variation of the perturbation $\tilde{v}(k_\perp)$. If 
$\tilde{v}(k_o)\sim -k_o\tilde{v}'(k_o)$ as obtained with a Gaussian function of characteristic spatial extension the atomic size 
$\sim 1/k_o$, we get a small negative phase-shift. If on the contrary, $\tilde{v}(k_o)$ is small compared to $k_o\tilde{v}'(k_o)$, 
then $\phi_o$ can take a significant value, which sign depends on the shape of $\tilde{v}$ around $k_o$. We see that the qualitative behavior of the density profile close to a singularity is expected to be 
universal, i.e. independent on the precise shape of $\tilde{v}$, only an additional phase-shift is model dependent.

\begin{figure}
 \begin{center}
 \includegraphics[width= 8cm]{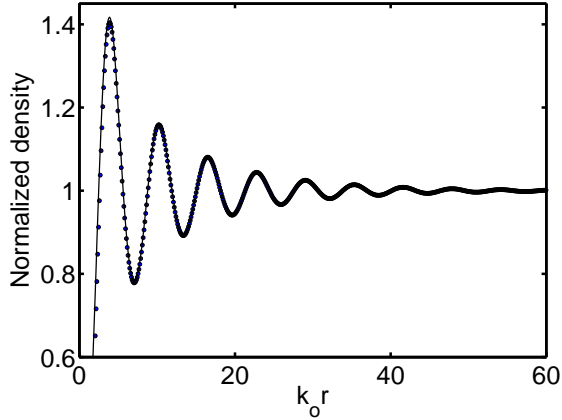}
 \end{center}
\caption{Comparison between the ``equilibrium'' density around a vortex (dots) in the Pomeau-Rica model, with the first order approximation Eq. \ref{linear} (full line). Eq. \ref{eq:BVproblem} is solved {\it via} a relaxation method. $k_1/k_o=0.0659$, $\phi=0.1694$.}
\label{vorprof}
\end{figure}

\section{Numerical solution of the Pomeau-Rica equation}
\label{numpr}

To compare the linear prediction obtained in Eq. \ref{foa} with the full rectilinear vortex solution of the Gross-Pitaevskii equation (Eq. 
\ref{ndpr}), we perform a numerical simulation aimed at evaluating the ground-state of Eq. \ref{ndpr} under the constraint that 
the phase turns of $2\pi$ along a line, as it was done in Refs. \cite{Stringari,Dal92,Berloff}. In this axisymmetric geometry, 
we are thus looking for an order parameter of the form $\psi(\vr,t)=f(r)e^{i\varphi}$, where $(r,\varphi,z)$ is the cylindrical 
coordinate system. The amplitude $f(r)$ is then given as the solution of the following boundary values problem
\begin{eqnarray}
\label{eq:BVproblem}
f''(r) + \frac{f'(r)}{r} +\left( \frac{4}{3}\pi\Lambda - \frac{1}{r^2}\right)f(r) 
 =\mbox{\hspace{20mm}}   \\
\Lambda f(r)\int_{r'=0}^{+\infty}\int_{\varphi=0}^{2\pi}\int_{z=-\infty}^{+\infty}
\theta\left( \sqrt{r^2 + r'^2 -2rr'\cos\varphi+z^2}\right)r'f^2(r')dr'd\varphi dz\nonumber
\end{eqnarray}
subject to the conditions $f(0) = 0$ and $f(+\infty)=1$.

Using the non-local potential $\theta$ of Pomeau and Rica \cite{Pomeau}, we can perform analytically the integration over $z$ and 
$\varphi$ using the Elliptic integrals of the first $\mathcal F(k) = \int_{0}^1 \frac{dt}{\sqrt{1-k^2t^2}\sqrt{1-t^2}}$ and 
second $\mathcal E(k) = \int_{0}^1 \frac{\sqrt{1-k^2t^2}}{\sqrt{1-t^2}}dt$ kind. 
We can show then that the RHS of Eq. \ref{eq:BVproblem} is 
given by, for $r\le 1$
\begin{eqnarray}
8f(r)\Lambda\int_{r'=0}^{1-r}\sqrt{1-(r-r')^2}\mathcal E\left(\sqrt{\frac{4rr'}{1-(r-r')^2}}\right)r' f^2(r')dr'  \nonumber \\ 
+8f(r)\Lambda \int_{1-r}^{1+r}\sqrt{rr'}\left[2\mathcal E\left(b\right)-(a+1)\mathcal F(b) \right]r' f^2(r')dr'   \nonumber
\end{eqnarray}
and for $r\ge 1$
$$ 8f(r)\Lambda \int_{r-1}^{r+1}\sqrt{rr'}\left[2\mathcal E\left(b\right)-(a+1)\mathcal F(b) \right]r' f^2(r')dr' \mbox{ ,}$$
where $a=\frac{r^2+r'^2-1}{2rr'}$, $b=\frac{\sqrt{2}}{2}\sqrt{1-a}$.
To numerically solve Eq. \ref{eq:BVproblem}, we use a relaxation method \cite{NumReci}. 
The Laplacian is furthermore estimated using a joint Crank-Nicolson and Gauss-Seidel algorithms \cite{NumReci} to ensure numerical 
stability. Starting with the uniform amplitude $f(r)=1$, 
the relaxation method converges towards the ``equilibrium'' distribution $f(r)$ that solves Eq. \ref{eq:BVproblem}.

In figure (\ref{vorprof}), we compare this approximation with the ``exact'' solution of the Pomeau-Rica model, 
for $\Lambda=40$, obtained with the above relaxation method. That is, we compare $f(r)^2$ with:
\begin{equation}
 \rho_{lin}=1+\frac{\sigma}{\sqrt{k_or}}\cos(k_or-\frac{\pi}{4}-\phi)\exp(-k_1r)
 \label{linear}
\end{equation}
$\sigma$ and $\phi$ being fitted for the best agreement. The parameter $\Lambda=40$, for which $k_o\approx 5.40$, is close to the ``spinodal'' value $\Lambda_s \simeq 43.43$, for which $\Omega=0$. As it can be seen in Fig. \ref{vorprof}, the first order approximation Equation \ref{linear} (solid line) almost perfectly fits the ``equilibrium'' density (dots). The value of $k_1/k_o$ is here $k_1/k_o=0.0659$. The value obtained for $\phi$ is $\phi=0.1694$, and $\sigma=-0.9917$.

\section{Melting pressure Helium}
\label{He}

In order to be closer to real superfluids, and to estimate to what extent the above ideas can apply to experimental systems, 
we consider now a more elaborate model. It has been used in a series of papers by Berloff and Roberts \cite{Berloff}, to which 
we refer for details. Apart from a larger number of parameters in tayloring the two-body interaction potential, 
it differs from the Gross-Pitaevskii approach by taking into account three-body interactions. Altogether, it allows to adapt 
the dispersion relation, both qualitatively and quantitatively, to that of the real superfluid of interest.

In this model, the Schr\"odinger equation \ref{pori} is changed to:
\begin{eqnarray}
 i\hbar\d_t\psi(\vr,t)=-\frac{\hbar^2}{2m}\Delta\psi(\vr,t) -\mu\psi(\vr,t) \mbox{\hspace{20mm}}  \nonumber \\
+ \psi(\vr,t)\left(\int d^3\vr'\,\cU(\vr-\vr')|\psi(\vr',t)|^2+W|\psi(\vr,t)|^4\right)
\label{nb}
\end{eqnarray}
where the three-body potential is treated as local, with intensity $W$.

Defining $a=\hbar/\sqrt{2m\mu}$, the same transformation than above, $\vr \rightarrow a\vr$, $\psi \rightarrow \sqrt{n}\psi$, $t \rightarrow (2ma^2/\hbar)t$, yields:
\begin{eqnarray}
 -i\d_t\psi(\vr,t)=\Delta\psi(\vr,t) + \psi(\vr,t)\left(1-\chi|\psi(\vr,t)|^4 \right) \nonumber \\
- \psi(\vr,t)\left(\int d^3\vr'\,U(|\vr-\vr'|)|\psi(\vr',t)|^2\right)
\label{ndnb}
\end{eqnarray}

The reduced potential $U$ is taken as (for convenience, we keep the same notations as \cite{Berloff}):
\begin{equation}
 U(r)=(\alpha+\beta A^2r^2+\delta A^4r^4)e^{-A^2r^2}+\eta e^{-B^2r^2}\mbox{ .}
\end{equation}
The dispersion relation is then given by:
\begin{equation}
 \omega^2=k^4+4k^2\chi+2k^2\tU(k)
\end{equation}
where $\tU(k)$ is the Fourier transform of $U(\vr)$.

The choice $A=1.3282$, $B=0.1992$, $\alpha=11.5881$, $\beta=-28.48$, $\delta=7.723$, $\eta=0.003775$, $\chi=3.5$, 
gives good agreement with the experimental dispersion curve at low temperature, at a pressure close to the melting one \cite{Bossy} (see figure \ref{disp}). In particular, we obtain good values not only of the roton minimum and its curvature, but also of the velocity of sound and the ``maxon'' energy.

\begin{figure}
 \begin{center}
 \includegraphics[width= 8cm]{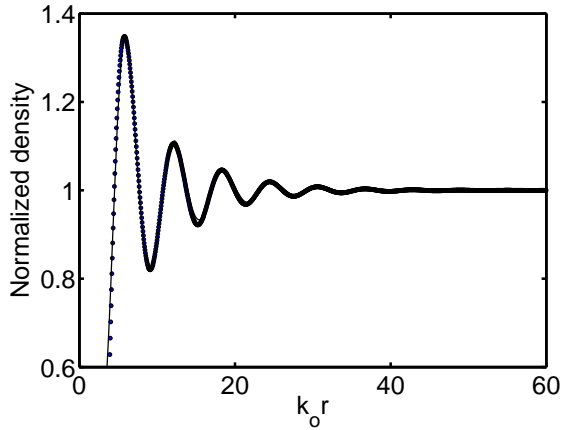}
 \end{center}
\caption{Comparison between the ``equilibrium'' density around a vortex (dots) in the Roberts model Eq. \ref{ndnb}, 
with the first order approximation Eq. \ref{foa} (full line). Eq. \ref{ndnb} is solved {\it via} a relaxation method. 
$k_1/k_o=0.1144$, $\phi= -0.99$.}
\label{vorprofnb}
\end{figure}

Similar numerical investigations of the density profile near a vortex as the ones done for the Pomeau-Rica case 
(Eq. \ref{eq:BVproblem}) can be performed. In this case, the RHS of Eq. \ref{eq:BVproblem}
can be analytically computed using Bessel functions (see Ref. \cite{Berloff} for details). 
The comparison between the ``equilibrium'' density and our first order approximation Eq. \ref{foa} is shown in figure 
\ref{vorprofnb}. Again, the agreement is very good, the discrepancy being visible only close to the second minimum. 
The value of $k_1/k_o$ is here $0.1144$. However, the value of $\phi$ is $\phi=-0.99$. Correspondingly, the hollow core of the vortex, that is the radius on which the density is nearly zero, is much wider with this model than with the Pomeau-Rica one. This quantitative phase-shift can be explained using a perturbation potential $\tilde{v}$ that is close to 0 around $k_o$.

\section{Comparison to variational calculations}
\label{reatto}

Up to now, we compared our linear approximation to Gross-Pitaevskii approaches. The structure of a vortex has also been studied {\it via} variational calculation approaches using appropriate trial wave functions \cite{Reatto2}. These calculations can be considered as numerical experiments on well controled fluids. In such a case, the authors have difficult access to the dynamical properties of their fluid model \cite{Reatto3}. It is then interesting to test both if the shape of the density structure is compatible with our predictions, and if the so extracted characteristics of the roton minimum match well the experimental measurements on the simulated state of the fluid.

\begin{figure}
 \begin{center}
 \includegraphics[width= 8cm]{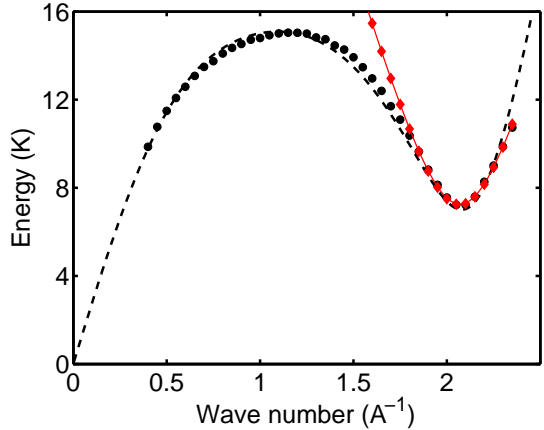}
 \end{center}
\caption{Filled black circles: fit of the experimental $^4$He dispersion curve (neutron scattering \cite{Bossy}). Filled red diamonds, and curve: parabolic fit, $k_1/k_o=0.12$, $k_o= 2.07\,10^{-10}$m$^{-1}$. Dashed line: Roberts' type model.}
\label{disp}
\end{figure}

\begin{figure}
 \begin{center}
 \includegraphics[width= 8cm]{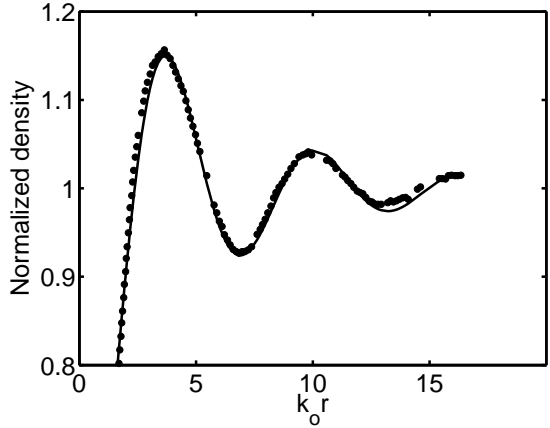}
 \end{center}
\caption{Comparison between the reference \cite{Reatto2} density profile (black dots, from their figure 3, dotted line), and our formula \ref{linear} with $k_1/k_o=0.12$, $\sigma=-0.4643$, $\phi=0.0186$.}
\label{Reat}
\end{figure}

In figure \ref{Reat}, we compare the results of reference \cite{Reatto2}, corresponding to $^4$He close to the melting pressure, to our formula for density variations close to a vortex line. The parameters are $k_o=2.07\,10^{10}$m$^{-1}$, and $k_1/k_o=0.12$, slightly different from those of the Roberts model treated above. However, the fit of the experimental \cite{Bossy} roton minimum with the formula:
\begin{equation}
 \omega(k)^2=c^2(k_1^2+(k-k_o)^2)
\end{equation}
where $c$ is here adjusted, is at least as good as with the Roberts' formula (see figure \ref{disp}, dashed line). Several informations can be drawn from the above results.
\begin{itemize}
\item First, the excitation energies corresponding to the trial wave function used in reference \cite{Reatto2} present a roton minimum whose wavelength and mass closely correspond to the experimental ones. This is a very good test of the pertinence of this wavefunction, which is difficult to obtain by other ways \cite{Reatto3}.

\item The phase shift is small $\phi=0.0186$. It is much smaller than obtained from a Roberts' model, even with a small three body term ($\phi=0.6650$). 

\item The amplitude of the density oscillations is smaller than with the corresponding Roberts' model.

\end{itemize}

Let us comment on the last two remarks. The wave function of reference \cite{Reatto2} has the peculiarity to give a finite value for the density on the vortex line, thus distributing the vorticity on a finite radius. No Gross-Pitaevskii model can do that. On the opposite, the Roberts type model tends to enlarge the hollow zone around the vortex line, especially the three body term. It naturally shifts  the oscillations outside, increasing $\phi$. It probably also enhances the perturbation induced by the vortex, and thus the amplitude of the oscillations.

This certainly raises a doubt about the utility of Gross-Pitaevskii approaches. As we have seen, the general shape of the oscillations, {\it i.e.} their pseudo-period and their amplitude decrease, is completely determined by the characteristics of the roton minimum. All the remaining physics reflects in the amplitude and the phase of these oscillations. Unfortunately, the Gross-Pitaevskii models give model dependent results, even if they reproduce the whole dispersion curve. We thus cannot trust the amplitude and phase predicted with these models.

However, it is interesting to look at the dependence of the amplitude {\it versus} $k_1/k_o$, {\it i.e.} the non dimensional roton gap. In reference \cite{Reatto2} the density profile around a vortex line has also been studied for low pressure $^4$He, where $k_1/k_o \simeq 0.16$. The obtained amplitude of the first peak is very close to the high pressure case (not shown). This almost insensitivity of the amplitude to the roton gap appears also with the simplest non local Gross-Pitaevskii model, the Pomeau-Rica one, except may be very close to the zero gap, where the amplitude seems to fall rapidly (however, the time necessary to reach the equilibrium shape diverges in this limit). This agreement between the two very different approaches gives us confidence in the near insensitivity of the amplitude with the roton gap, which will be important in the next section.

\section{Energy considerations}
\label{eng}

Even within our linear approximation of the density deformations, it is possible to have a look at their contribution to the vortex energy. It comes from the fact that, in a free propagating linear excitation, the kinetic energy and the potential deformation energy are equal in average (this deformation energy not only comes from the two body interaction potential between atoms, but also from the real part of the wavefunction gradient, the imaginary part giving the kinetic energy).  Furthermore, mass conservation allows to relate the deformation amplitude of this excitation to the longitudinal velocity (i.e. $u(k)$ the component along $\k$ of the Fourier transform of velocity):
\begin{equation}
\partial_t\rho + \partial_j\rho u_j=0 \Longrightarrow  -\omega(k)\rtdr(k) +ku(k)=0 
\label{mass}
\end{equation}

Thus the contribution of a wave vector $\vec{k}$ to the deformation energy is:
\begin{equation}
 e(k)=\frac{\rho_o}{2}\frac{\omega(k)^2}{k^2}\left[\rtdr(k)\right]^2\mbox{ ,}
\end{equation}
and the total contribution of the oscillations to the energy per unit length of the vortex is then:
\begin{equation}\label{eq:enerdeform}
\epsilon  =  \frac{1}{(2\pi)^3}\int e(k)\,d^3\vec{k}\mbox{ .}
\end{equation}

We are interested here in quantifying the energy induced by the oscillations as given by the simplified form of the perturbation Eq. \ref{tdr4}. In particular, we are not taking into account the energy deformation induced by the vortex core, this can be done by other means \cite{Stringari}. In this spirit, we restrict the domain of integration in Eq. \ref{eq:enerdeform} to a neighborhood of the roton gap $k_0$. As we saw, this approximation is consistent when the roton gap is deep, i.e. when $k_1=\Omega/c \ll k_0$ (see Eqs. \ref{foa} and \ref{foa2}). For example, the integration domain can be restricted to $k_0 \pm nk_1$, with $n$ an integer such that $nk_1\ll k_0$, and we get, using the approximated form of $\rtdr(k)$ (Eq.  \ref{tdr4})
\begin{equation}\label{eq:1stapproxener}
\epsilon \approx \frac{1}{4\pi}\frac{\rho_o}{\Omega^2}\int_{k_{\perp}=k_0-nk_1}^{k_0+nk_1}\frac{dk_{\perp}}{k_{\perp}}\frac{\left[  \tilde{v}(k_o) + \tilde{v}'(k_{o}) (k_{\perp}-k_o) \right]^2}{1+\left(\frac{k_{\perp}-k_0}{k_1}\right)^2}\mbox{ .}
\end{equation}
In the integrand of the former integral (Eq. \ref{eq:1stapproxener}), the numerator can be expanded and gives three terms, two of them being proportional to $(k_{\perp}-k_o)$ and $(k_{\perp}-k_o)^2$. Their contribution to the energy (Eq. \ref{eq:1stapproxener}) are of the order of $n(k_1/k_0)^3$ and, henceforth, will be neglected. It also shows that, in the limit of vanishing roton gap $k_1/k_0\rightarrow 0$,  the phase-shift $\phi_o$ contributes with a negligible factor to the energy. The remaining leading term, proportional to $\tilde{v}(k_o)^2\approx \gamma^2$, $\gamma^2$ being itself proportional to $\sigma^2$ (see Eq.  \ref{foa2}), can be shown to be of the order of $(k_1/k_0)$ and independent on $n$. We finally get for the deformation energy
\begin{equation}
\epsilon \approx \frac{\pi\rho_oc^2\sigma^2k_1}{2k_o^3}\mbox{ .}
\label{energy}
\end{equation}

Looking at the expression of $\epsilon$ (Eq. \ref{energy}), we see that it contains $k_1$ at the numerator. Let us remark that we checked numerically on the Pomeau-Rica model that the amplitude $\sigma^2$ does not diverge when the parameter $\Lambda$ tends to the spinodal value $\Lambda_s\approx 43.43$ (data not shown). It means that, while the range of the oscillations increase when $k_1$ goes to zero, at the same time their total energy goes to zero. It is understandable: this is because their energy goes to zero that their range can increase. On the one hand, it suggests that these oscillations poorly contribute to the total energy per unit length of the vortex, which should be dominated by the kinetic energy. On the other hand, if fully attributed to the kinetic energy, the above decrease of the oscillations energy will be interpreted as an increase of the hollow core radius.

This paradox was underlined in \cite{Reatto2}. Indeed, Equation \ref{energy}, applied to liquid helium at the freezing pressure, gives a deformation energy of 0.0176K/\AA, {\it i.e.} within the error bar of the vortex energy (1.90(3)K/\AA\ \cite{Reatto2}). A precise determination of this deformation energy would give an estimate of the roton gap, thus completing all the informations on the roton dispersion curve.

\section{Discussion and Conclusion}
\label{conc}

We have shown that the density profile of the superfluid close to a singularity is expected to be universal when the roton gap $\hbar\Omega$ is small. We have derived an explicit form of this behavior (Eq. \ref{foa}) valid as well for real superfluids as for Gross-Pitaevskii models. We have emphasized that only the additional phase-shift $\phi_o$ and the overall amplitude $\sigma$ depend on the precise shape of the model. The linear prediction compares as well with numerical simulation of the ground state of the Pomeau-Rica and Berloff-Roberts models, as with first principle calculations in presence of a vortex. However, while in the first case, the order parameter is concerned in a Gross-Pitaevskii approach, in the second case, the oscillations directly concern the fluid density.

This agreement allows to use the oscillations in a reciprocal way, and to deduce some characteristics of the roton minimum from their shape. This is particularly useful with the first principle calculations where a direct ``measurement'' of the dynamic properties of the obtained ground state is very difficult \cite{Reatto3}. The roton wave number, and the width of the roton minimum can be deduced from the wavelength and the decrement of density oscillations close to a singularity. Even the roton gap could be deduced from a precise determination of the energy of these oscillations.

As for the vortex, the energy of the corresponding density oscillations is negligible compared to the kinetic energy associated with the fluid circulation around it. As a consequence, the vortex dynamics is unaffected by these density oscillations as far as the curvature of the vortex is smaller than $k_1$. In particular, small amplitude Kelvin waves should be unaffected \cite{Kelvin}. However, large amplitude Kelvin waves, with wavelength smaller than $1/k_1$ will be affected. This point goes out of the scope of this paper.

While we focused on $^4$He close to the freezing pressure, our conclusions can be extended to other examples. Metastable overpressurized Helium has already been obtained within nanopores \cite{Bossy2}, or transiently, with acoustic oscillations \cite{Werner}. Many attempts aimed at obtaining metastable superfluid molecular hydrogen \cite{Maris}, and estimations give good hope for a next future \cite{Ceperley}. Finally, even cold atomic gases seem able to sustain roton like excitations \cite{Santos}. In all these cases, the dynamic properties of rotons can be precisely deduced from static measurements.

\begin{acknowledgements} 
Thanks are due to J. Bossy for an interesting discussion and for providing us with unpublished data, and to D. Galli and L. Reatto for additional data and discussions. This work has been supported by the Agence Nationale pour la Recherche, under the contract ANR-09-BLAN-0094-01.
\end{acknowledgements}




\end{document}